\documentclass[review,authoryear]{elsarticle}

\usepackage[utf8]{inputenc}
\usepackage{longtable,booktabs}
\usepackage{threeparttable} 
\usepackage{threeparttablex}
\usepackage{dcolumn}
\usepackage[margin=3cm]{geometry}
\usepackage{lineno,hyperref}
\usepackage{subfig}
\usepackage{rotating}
\modulolinenumbers[5]

\graphicspath{{FiguresDraft/}}

\usepackage{array}
\usepackage{calc} 
\usepackage{amsmath,amssymb,latexsym} 
\usepackage{enumerate} 
\usepackage[dvipsnames]{xcolor}
\usepackage{amsmath,amssymb,latexsym} 
\usepackage{enumitem} 
\usepackage{xcolor,colortbl}

\biboptions{numbers,sort&compress}

\journal{???}

\DeclareMathOperator{\Unif}{Unif}

\DeclareMathOperator{\upi}{\underline\pi} 

\begin{document}

\begin{frontmatter}

\title{New algorithms and goodness-of-fit diagnostics from\\remarkable properties of ranking models}


\author[mymainaddress]{Cristina Mollica\corref{mycorrespondingauthor}}
\cortext[mycorrespondingauthor]{Corresponding author}
\ead{cristina.mollica@uniroma1.it}

\author[mysecondaryaddress]{Luca Tardella}

\address[mymainaddress]{Dipartimento di Metodi e Modelli per l'Economia, il Territorio e la
  Finanza, Sapienza Universit\`{a} di Roma}
\address[mysecondaryaddress]{Dipartimento di Scienze Statistiche Sapienza Universit\`{a} di Roma}

\begin{abstract}
The forward order assumption 
postulates that the ranking process of the items is carried out by sequentially assigning the positions
from the top (most-liked) to the bottom (least-liked) alternative. This assumption has been recently relaxed with the \textit{Extended Plackett-Luce model} (EPL) through the introduction of the discrete \textit{reference order} parameter, describing the rank attribution path. By starting from two formal 
properties of the EPL, the former related to the inverse ordering of the item probabilities at the first and last stage of the ranking process and the latter well-known as \textit{independence of irrelevant alternatives} (or \textit{Luce's choice axiom}), we derive novel diagnostic tools for testing the appropriateness of the EPL assumption as the actual sampling distribution of the observed rankings.
Besides contributing to fill the gap of goodness-of-fit methods for the family of multistage models, we also show how one of the two statistics can be conveniently exploited to construct a heuristic method, that surrogates the maximum likelihood approach for inferring the underlying reference order parameter. 
The relative performance of the proposals compared with more conventional approaches is illustrated by means of extensive simulation studies.


\end{abstract}


\end{frontmatter}


\section{Introduction}
\label{s:intro}
%
Let us consider an experiment in which a sample of $N$ judges is asked to rank a set $I=\{1,\dots,K\}$ of $K$ labeled alternatives, namely \textit{items}, according to a certain criterion. The final outcome of the comparative evaluation is an ordered sequence collecting the positions attributed to each object, called ranking. Formally, a \textit{ranking} is a vector $\pi=(\pi(1),\dots,\pi(K))$
where the entry $\pi(i)$ indicates the position attributed to the $i$-th alternative. Equivalently, data can be recorded in the \textit{ordering} format $\pi^{-1}=(\pi^{-1}(1),\dots,\pi^{-1}(K))$, where the generic component $\pi^{-1}(j)$ indicates the item ranked in the $j$-th position. 
This implies that ranking/ordering data are multivariate ordinal data taking values in the set of permutations $\mathcal{S}_K$ of the first $K$ integers.

These observations are typical in several areas of research, involving market surveys on preferences for consumer goods \citep{Yao,Gormley:Murphy2010}, psychological/behavioral studies on attitudes \citep{Croon89,Yu2005} and voting systems allowing for the elicitation of an ordering of the candidates in the ballots \citep{Gormley:Murphy-AnnalsApplied,Lee:Yu2012}. Another relevant field concerns the competition/sport context, where the competitors (players, teams) are ranked according to a certain measure of ability, such as the arrival time in a race or the score collected during a championship, see \cite{Henery-Royal,Stern1990,Henderson}. 

The broad statistical literature on methods and models for analysing ranking data is reviewed in \cite{Marden} and, more recently, in \cite{Alvo} and \cite{Crispino}. 
By considering the parametric modelling approach, distributions over the permutation set are traditionally classified into four main categories: (i) order statistics models (OS), whose seminal work is represented by \cite{Thurstone}; (ii) paired comparison models \citep{Bradley76,Bradley84}; (iii) distance-based models (DB), see \cite{Mallows} and \cite{Fligner:Verducci-Royal} and (iv) stagewise models \citep{Fligner:Verducci-American}.
This work concentrates on the parametric family (iv), relying on the idea that the ranking process can be decomposed into consecutive stages
for each position that has to be assigned, in particular on the 
\textit{Extended Plackett-Luce model} (EPL) introduced by \cite{Mollica:Tardella}. The EPL generalizes the popular \textit{Plackett-Luce} model (PL), presented by \cite{Luce} and \cite{Plackett1968}, by relaxing the implicit \textit{forward order} assumption, according to which the ranking process of the alternatives proceeds sequentially from the most-liked to the least-liked item. The extension was accomplished by adding the \textit{reference order parameter} $\rho= (\rho(1),...,\rho(K))$ in the PL
formulation. It
indicates the rank assignment order, that is, $\rho(t)$ denotes the position attributed at the $t$-th stage.

One aspect which is very often overlooked
in the ranking literature concerns the assessment of model adequacy to the observed data.
Traditional approaches are based on the construction of diagnostics to detect possible lack-of-fit of generic sample quantities, rather than to asssess the conformity of the data with peculiar features of the postulated model. Moreover, the investigation of the effectiveness of these methods has been limited to few parametric families, such as for example the DB \citep{Feigin:Cohen,Cohen:Mallows} and the OS \citep{Yao,Yu,Tsai2000}.
This has motivated us to the develop
some original 
tools to appropriately 
check the model mis-specification issue for the class of multistage models, specifically
for the EPL assumption as the data generating mechanism.
We first
introduced
two novel test statistics:
the former is based on a formal property of the EPL class which has not been considered and proven earlier in the literature, whereas the latter relies on the well-known assumption of the PL distribution, known as \textit{independence of irrelevant alternatives} (or \textit{Luce's choice axiom}). Through an extensive simulation study under different model specifications, we compared the power of the two diagnostics with that of more frequently-used test statistics. The analysis of simulated data revealed
the relative merits and limits of the competing diagnostics for the EPL assumption.
Secondly, we reconsider the former EPL property from the inferential perspective as the key element of an heuristic method to estimate $\rho$. We implemented a simulation study to quantify the inferential ability of the proposed likelihood-free estimation strategy to recover the actual reference order parameter. It showed a promising and consistent behavior of the heuristic technique, that could be exploited to reduce the computation burden affecting the EPL estimation task.


The outline of the article is the following. In Section \ref{s:pl}, the EPL and the related inferential approaches are briefly recalled. Section \ref{s:moddiag} provides a detailed review of the strategies proposed in the literature to address the model assessment issue for ranking distributions within the frequentist and the Bayesian paradigm. Two novel goodness-of-fit diagnostics for the EPL parametric class are then introduced in Section \ref{s:newdiag} and a comparative evaluation with more standard measures of model adequacy follows in Section \ref{s:comp}. An original heuristic method to infer the reference order parameter from the maximum likelihood-free perspective is defined in Section \ref{s:euristic} and its
effectiveness is investigated
with an extensive simulation study.
Concluding remarks and proposals for future work are discussed in Section \ref{s:conc}.


\section{The Extended Plackett-Luce model}
\label{s:pl}


\subsection{Model specification and inference}
\label{ss:modspec}
%

In order to explore alternative meaningful rank assignment orders for the choice process and increase the flexibility of the PL distribution,
\cite{Mollica:Tardella} suggested the PL extension based on the relation of the canonical forward order assumption. 
The probability of a generic ordering under the EPL assumption can be written as
\begin{equation}
\label{e:EPL}
\mathbf{P}_\text{EPL}(\pi^{-1}|\rho,\underline{p})=\mathbf{P}_\text{PL}(\pi^{-1}\circ\rho|\underline{p})
=\prod_{t=1}^K\frac{p_{\pi^{-1}(\rho(t))}}{\sum_{v=t}^Kp_{\pi^{-1}(\rho(v))}}\qquad\pi^{-1}\in\mathcal{S}_K,
\end{equation}
where the symbol $\circ$ denotes the composition between two permutations, the reference order $\rho=(\rho(1),\dots,\rho(K))$ is a discrete parameter, specifically a permutation of the first $K$ integers, and the positive quantities $p_i$'s are referred to as \textit{support parameters}. These are
proportional to the probabilities for each item
to be ranked
in the position
indicated by the first entry of $\rho$. Hereinafter, we will shortly refer to model~\eqref{e:EPL} as $\text{EPL}(\rho,\underline{p})$.
Inference on the EPL and its generalization into a finite mixture framework was originally addressed from the Maximum Likelihood Estimation (MLE) perspective in \cite{Mollica:Tardella} via the hybrid Expectation-Maximization-Minorization (EMM) algorithm. Recently, \cite{MollicaSMA} introduced the Bayesian version of the EPL where, for the prior specification,
independence of $\rho$ and $\underline{p}$ was assumed together with
a discrete uniform distribution on $\mathcal{S}_K$ for the reference order and 
conjugate Gamma densities for the support parameters. The data augmentation with
the latent quantitative variables
underlying the observed ordinal sequences 
contributed to facilitate the Bayesian estimation task from an analytical point of view, see \cite{MollicaPSY} for more in-depth details. 
A tuned joint Metropolis-within-Gibbs sampling (TJM-within-GS) was developed to conduct approximate posterior inference on the mixed-type parameter space. The TJM-within-GS was also adapted for the inference of the Bayesian mixture of EPL distributions \cite{MollicaCassino} and for the EPL with order constrains on the reference order \cite{MollicaPalermo}.

\section{Goodness-of-fit diagnostics for ranking models: a review}
\label{s:moddiag}

\subsection{Frequentist literature}
\label{s:moddiagfreq}
In the frequentist framework, the problem of assessing model adequacy to the observed data can be addressed with methods such as the likelihood ratio or the Chi-squared test. One of the first contributions for the evaluation of the fit of ranking models was introduced by \cite{Cohen:Mallows}, that handled separately the two cases $K<5$ and $K\geq 5$. In the former situation, the cardinality of the ranking space is manageable, so that each of the $K!$ ranked sequences can be
regarded as a single category of a multinomial distribution. Thus, when $K$ is small and $N$ is large enough, model fit can be assessed by quantifying the difference between the observed frequencies of each ranking and the expected ones under the estimated model. 

However, the rapidly increasing size of the ranking space makes this approach impracticable for larger values of $K$, due to the possible occurrence of sparse data. In fact, null or low frequencies encountered for some ranking patterns imply that asymptotic properties of the aforementioned test statistics no longer hold. So, for the case $K\geq 5$, a more parsimonious representation of the data is typically needed and \cite{Cohen:Mallows} suggested to identify relevant partitions of the permutation set capturing meaningful features of the preference elicitation. In so doing, model fitness can be then evaluated on each subset as previously described for smaller values of $K$. Some examples are the groupings of the rankings according to their Kendall distance from the estimated modal sequence, formerly proposed by \cite{Feigin:Cohen} as a natural diagnostic tool for the $\phi$ model, that is, the DB model with the Kendall distance originally presented by \cite{Mallows}. Additionally, \cite{Cohen:Mallows} adopted the partition induced by the pairwise comparisons (PC), counting the number of times that item $i$ is ranked higher than item $i'$ ($1\leq i<i'\leq K$), and employed it to check the adequacy of  the Thurstone-Mosteller-Daniels model, that is, the OS model with independent normal latent scores introduced by \cite{Daniels}. For the latter, under the independence assumption, they computed the standardized deviates $Z_{ii'}$ to measure the difference between expected and observed frequencies and also displayed the absolute values of the statistics into a half normal probability plot, to better highlight 
local misfits of the data. 
A similar method was employed by \cite{Yu}, who divided the rankings into the subgroups of sequences with the same item in the top position and compared the sample top frequencies with those expected under the OS model with correlated normal utilities.
\cite{Maydeu2005} have proposed to overcome the problem of sparse data by applying the adjusted test statistics defined in \cite{Satorra1994,Satorra2001}, in order to check the adequacy of unrestricted OS models.

\subsection{Bayesian literature}
\label{s:moddiagbay}
Relevant goodness-of-fit diagnostics came up also  from more recent works of the Bayesian ranking literature. 
In this framework, goodness-of-fit assessment accounts for the randomness of the parameter, 
rather than relying on a single point estimation as typical in the frequentist domain. Specifically, the Bayesian approach 
relies on the construction of a \textit{discrepancy variable}, that depends on both data and parameters and can be employed in the so-called \textit{posterior predictive check}. The core idea is to assess the conformity of the observed value of the discrepancy in the realized sample with its predicted distribution under the estimated model.
The computation of the reference distribution of the discrepancy measure under the assumed model is straightforward when a sample from the posterior distribution is available, as the output of MCMC methods. See \cite{Meng} and \cite{Gelman:Meng:Stern} for a general description of Bayesian assessment methods via posterior predictive checks. 

For model-based ranking analysis,
\cite{Yao} focussed on paired, triple and quadruple comparisons as empirical summaries. \cite{Tsai2000} conducted an extensive Monte Carlo simulation study to evaluate the validity of the posterior predictive check for testing the adequacy of alternative OS models. They considered different discrepancy measures and analyzed the effect of the number $K$ of alternatives, the sample size $N$ and the type of mis-specification on the lack-of-fit detection. In particular, they proved the usefulness of the marginal rank distributions, also known as \textit{first-order marginals}, providing the counts that each item $i$ is ranked in position $j$. Finally, \cite{MollicaPSY} constructed two discrepancy variables, based respectively on the top and the PC frequencies, and described how to use them to conduct the posterior predictive check for the Bayesian PL mixture unconditionally and conditionally on the length of the observed partial top rankings.

\section{Novel EPL diagnostics}
\label{s:newdiag}

The reviews provided in Section \ref{s:moddiag} reveal that specific diagnostic tools to evaluate model adequacy of the class of multistage ranking distributions are very limited 
and their effectiveness has not been deeply explored. One of the objectives of the present work is a contribution to address the goodness-of-fit issue for the EPL specification.


\subsection{Testing the property of inverse monotonicity of the last-stage item probabilities}
\label{ss:eplprop}

Let us suppose that we have some data simulated from an $\text{EPL}(\rho,\underline{p})$. We expect the marginal frequencies of the items at the first stage to be ranked according to the order of the corresponding support parameter component. On the other hand, 
we expect the marginal frequencies of the items at the last stage to be ranked according to the reverse order of the corresponding support parameter component. This claim relies on the ordering relations linking the marginal item probabilities concerning, respectively, the first and the last stage of the ranking process. The formal proof is provided in Appendix A. This is based on an appropriate method to index the sequences contributing to the construction of the marginal item probabilities, that facilitates the ordinal comparison between them. Hereinafter, we will refer to this property as \textit{inverse monotonicity of the last-stage item probabilities}.
One can then derive that the ranking of the marginal frequencies of the items corresponding to the first and last stage should sum up to $(K+1)$, no matter what their support is. Of course, this is less likely to happen when the sample size is small or when the support parameters are not so different of each other. In any case, one can define a test statistic by considering,
for each couple of integers $(j,j')$ candidate to represent the first and the last stage ranks, namely $\rho(1)$ and $\rho(K)$, 
a discrepancy measure $T_{jj'}(\upi)$
between $K+1$ (the sum of the expected ranks) and the
sum of the observed ranks of the
frequencies corresponding to the same item extracted in the first and in the last stage. Formally, let $\underline{r}^{[1]}_j=(r^{[1]}_{j1},\dots,r^{[1]}_{jK})$ and 
$\underline{r}^{[K]}_{j'}=(r^{[K]}_{j'1},\dots,r^{[K]}_{j'K})$ be
the marginal item frequency distributions for the $j$-th  and $j'$-th positions, 
to be assigned respectively at the first [1] and last [K] stage.
In other words, the generic entry $r^{[s]}_{ji}$ is the number of times that item
$i$ is ranked $j$-th at the $s$-th stage. The proposed EPL diagnostic relies on the following discrepancy
\begin{equation}
\label{t:Tentry}
T_{jj'}(\upi)=\sum_{i=1}^K\lvert\text{rank}(\underline{r}^{[1]}_{j})_i+\text{rank}(\underline{r}^{[K]}_{j'})_i-(K+1)\rvert,
\end{equation}
implying that the 
smaller the value $T_{jj'}(\upi)$, the larger the plausibility that the two integers $(j,j')$ represent the first and the last components of the reference order. In this sense, $T_{jj'}(\upi)$ is a measure of the closeness of the positions $j$ and $j'$ in the rank attribution path.
To globally assess the conformity of the sample with the EPL, we consider the statistic
%
\begin{equation}
\label{t:T}
T_{m}(\upi)=\underset{j<j'}{\text{min}}\,T_{jj'}(\upi).
\end{equation}
%

\subsection{Testing the property of independence of irrelevant alternatives}
\label{ss:iiaprop}
%
With the aim at further enlarging the collection of diagnostics of fit for the EPL class, we focus our attention also on another specific property of the PL. In particular, we consider the distinguishing assumption of the PL knows as \textit{Luce's choice axiom} or \textit{independence of irrelevant alternatives} (IIA) to construct a further specification test. The IIA states that the relative preferences between two items $i$ and $i'$ does not depend on the liking for the other alternatives belonging to the choice set. 
For the EPL, the IIA hypothesis implies that the probability ratio of selecting item $i$ over item $i'$ is constant over the stages of the ranking process  (constant ratio rule), as long as the two items are both still available. Formally, let $I_{st}=I\setminus\{\pi_s^{-1}(\rho(1)),\dots,\pi_s^{-1}(\rho(t-1))\}$ be the choice set composed of the alternatives available at the $t$-th stage for unit $s$, i.e., those items which have not been selected by the ranker $s$, and hence removed from the comparison, before stage $t$. By introducing the binary indicator
\begin{equation*}
\xi_{ii'st}
=\begin{cases}
      1\qquad \text{$i,i'\in I_{st}$}, \\
      0\qquad \text{otherwise},
\end{cases}
\end{equation*}
one can compute the observed PC at stage $t$ where item $i$ is selected before item $i'$ as 
$$\tau_{ii't}=\sum_{s=1}^N\xi_{ii'st}I_{[\rho^{-1}(\pi_s(i))<\rho^{-1}(\pi_s(i'))]}.$$
The IIA implies that the expected PC frequency at stage $t$ of choosing item $i$ over item $i'$ is
$$\tau^*_{ii't}=N_{ii't}\frac{p_i}{p_i+p_{i'}},$$
equal to the product between the total number $N_{ii't}$ of PCs between $i$ and $i'$ at stage $t$, given by
$$N_{ii't}=\tau_{ii't}+\tau_{i'it}=\sum_{s=1}^N\xi_{ii',st},$$ and the theoretical PC probability under the EPL, concerning the choice of the two items from the entire set $I$ of the $K$ alternatives.
Hence, a chi-squared statistic for the IIA assumption can be defined as follows
\begin{equation}
\label{e:IIA}
X^2_{\text{IIA}}=\sum_{t=1}^{K-1}\sum_{i<i'}\frac{(\tau_{ii't}-\tau^*_{ii't})^2}{\tau^*_{ii't}}.
\end{equation}
The IIA diagnostic operates in a stagewise manner by assessing the relative selection probability of each pair of items $(i,i')$ at each stage $t=1,\dots,K-1$ of the ranking process. 

%
%
%
%
%
%
%
%
%
%
\section{Comparative assessment of goodness-of-fit diagnostics for ranking models}
\label{s:comp}
%

After introducing novel test statistics, one should enquire into their power, for instance through a bootstrap approach, and preferably compare it with that of some standard goodness-of-fit tools for ranking models. To this aim, we conducted a simulation study under alternative model specifications, involving the comparison with the chi-squared statistics based on the top frequencies, the PC and the first-order marginals, given respectively by
\begin{equation*}
X^2_{\text{TOP}}=\sum_{i=1}^K\frac{(m_{1i}-m^*_{1i})^2}{m^*_{1i}}\qquad
X^2_{\text{PC}}=\sum_{i<i'}\frac{(\tau_{ii'}-\tau^*_{ii'})^2}{\tau^*_{ii'}}\qquad
X^2_{\text{M}}=\sum_{j=1}^K\sum_{i=1}^K\frac{(m_{ji}-m^*_{ji})^2}{m^*_{ji}},
\end{equation*}
where the expected frequencies are $m^*_{1i}=Np_i$, $\tau^*_{ii'}=N\frac{p_i}{p_i+p_{i'}}$, whereas $m^*_{ji}$ were estimated with a Monte Carlo simulation 
Note that $X^2_{\text{M}}$ is a stagewise extension of the classical chi-squared statistic $X^2_{\text{TOP}}$. In fact, the latter is obtained from $X^2_{\text{M}}$ by considering only the term $t=1$ of the outer sum, concerning the marginal item distribution in the first position, that is 
$$m_{1i}=\sum_{s=1}^NI_{[\pi_s(i)=1]}.$$ 
Similarly, the IIA diagnostic can be regarded as a stagewise generalization of $X^2_{\text{PC}}$.
For the latter, the comparison between item $i$ and $i'$ is considered only at the first stage, that is, in the context of the whole item set $I$ for which $N_{ii'1}=N$. 

A comparative evaluation of the model specification tools described in Section \ref{s:newdiag} was carried out by means of an extensive simulation study. For each possible combination $(K,N)$, with values varying respectively in the grids $K \in \{5,10,15\}$ and $N \in \{300,450,600\}$, we drew $100$ datasets with $N$ orderings of $K$ items from the following ranking distributions:
\begin{enumerate}
      \item EPL;
      \item DB with the Kendall distance (DB-Kend);
      \item DB with the Cayley distance (DB-Cay);
      \item DB with the Hamming distance (DB-Ham);
      \item TH with normal latent scores (TH-norm); 
\end{enumerate}
where the true parameter values were uniformly generated. To approximate the probability that, under the EPL assumption, the theoretical distribution of the test statistic is greater than or equal to the observed value, we employed the $p$-values obtained from the bootstrap method by  considering 1000 datasets drawn from the inferred EPL. Deviations from the EPL model should yield greater values of the test statistics and, hence, smaller $p$-values. Finally, for each model adequacy criterion, we estimated the mis- and correct rejection rates with the relative frequency of the times that the $p$-value was smaller than or equal to the conventional 0.05 critical threshold. 

\begin{figure}
\centering
{\includegraphics[scale=0.285]{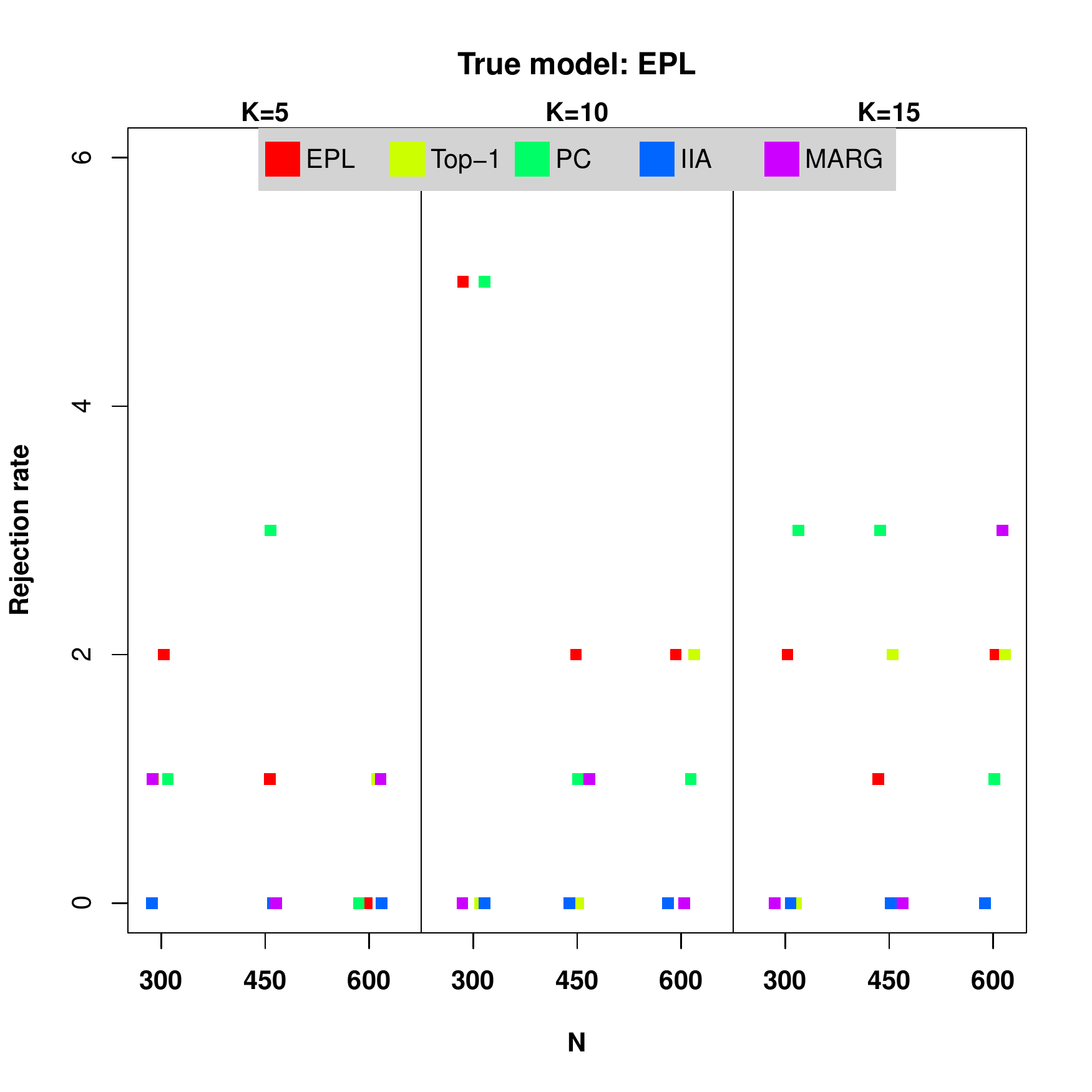}}
{\includegraphics[scale=0.285]{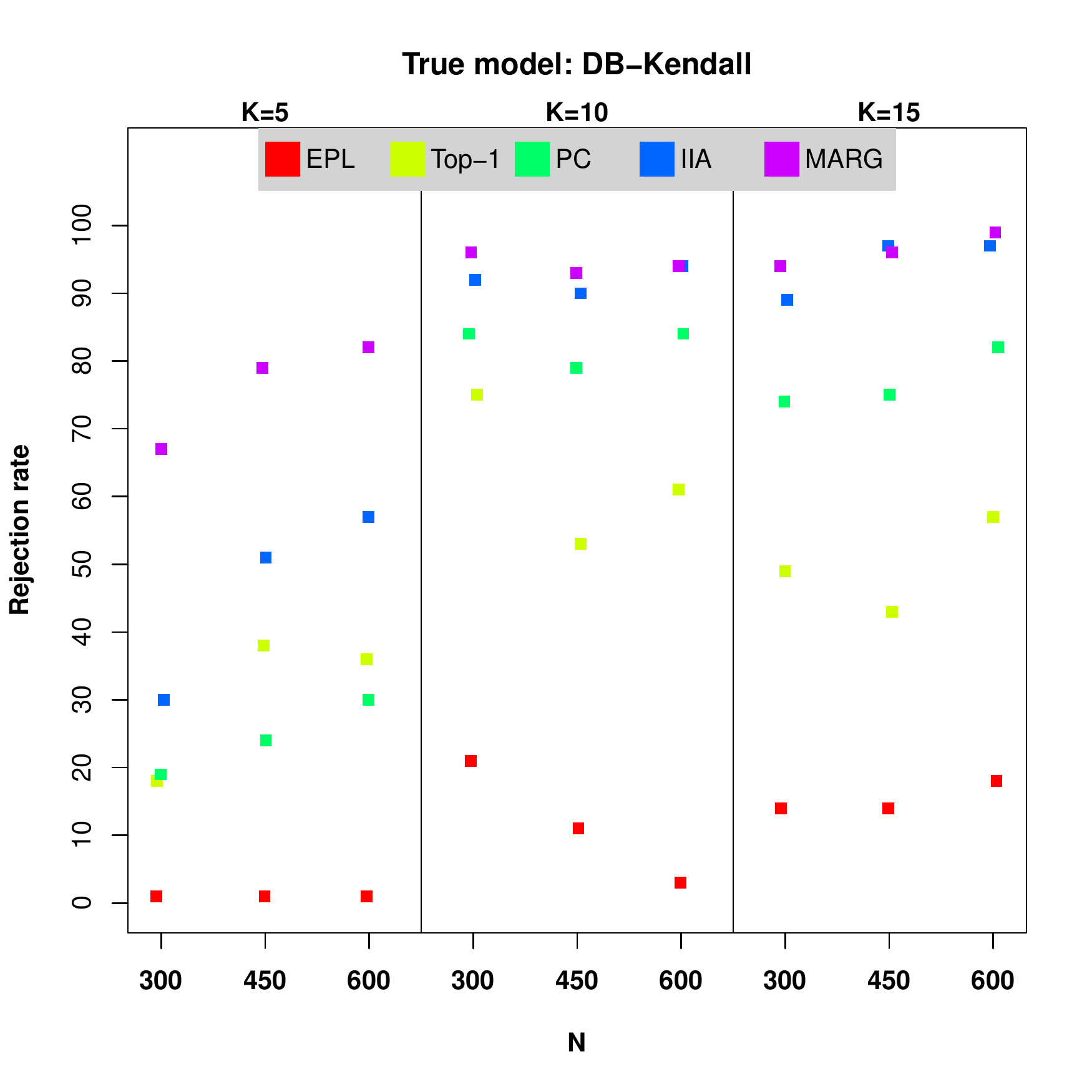}}
{\includegraphics[scale=0.285]{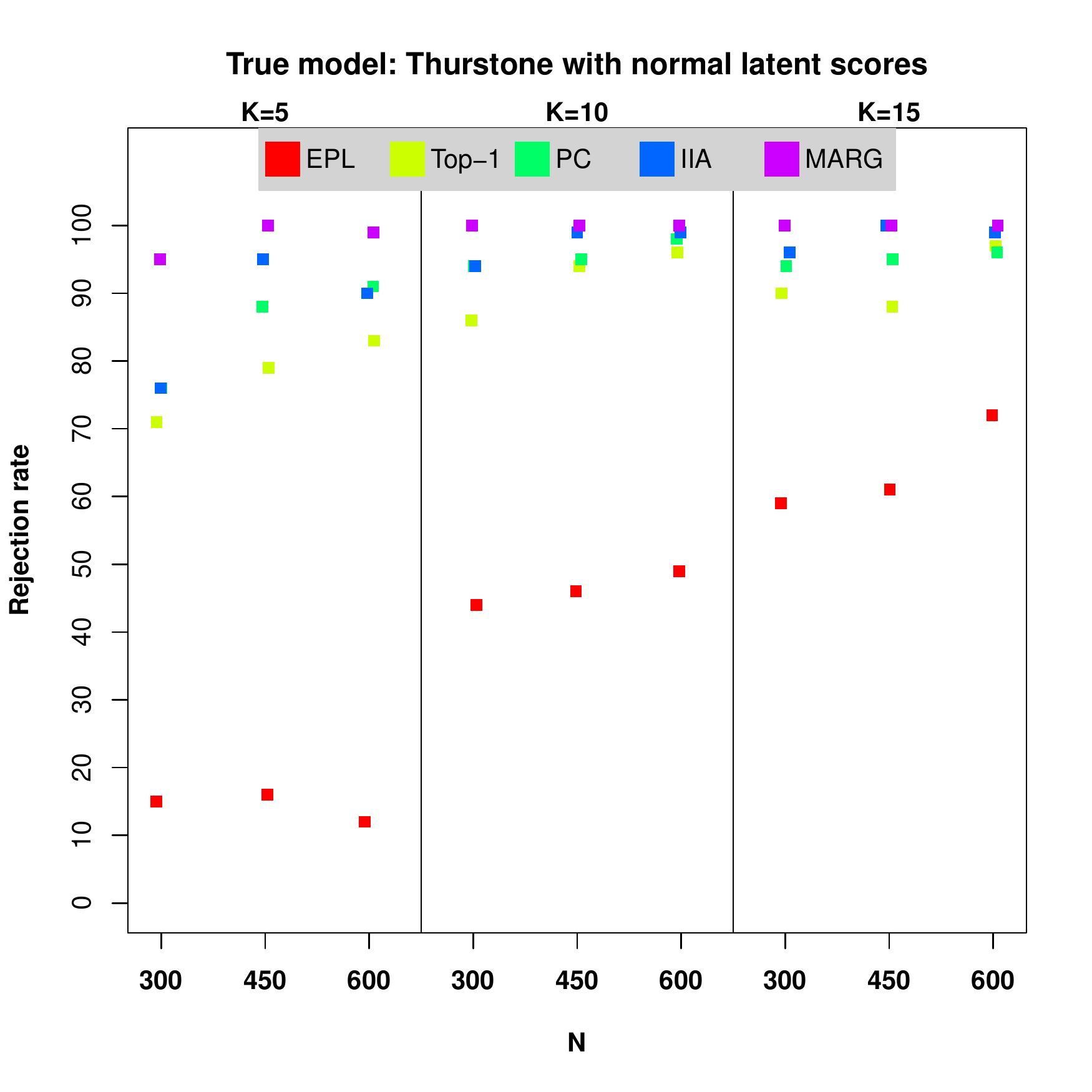}}
\caption{Rejection rates of the EPL assumption for alternative goodness-of-fit diagnostics computed on simulated data from different model scenarios. The reference distributions of the test statistics under the EPL assumption have been approximated with the bootstrap method.}
\label{fig:rejectrates}
\end{figure}

The simulation study revealed a satisfactory performance of all the considered diagnostics regarding the rates of mis-rejections, testified by estimated Type I error probabilities below 0.05 (Figure \ref{fig:rejectrates}, left). On the other hand, noteworthy differences emerged in terms of the power. Firstly, measure \eqref{t:T} exhibited a consistent poor behavior of the estimated power under each considered model scenario. For instance, see the rates of correct rejections under the two population scenarios shown in Figure \ref{fig:rejectrates} (center and right).
At least two motivations can be put forward to argue this evidence. The former is related to the formal definition of $T_{m}(\upi)$; in fact, \eqref{t:T} is a parameter-free measure based on the ranks of the expected marginal frequencies, rather than on the computation of the parameter-dependent first- and last-stage theoretical probabilities. This makes $T_{m}(\upi)$ by construction a rougher diagnostic in the comparison with the other statistics.
Secondly, the remarkably low power of \eqref{t:T} under the DB with the Kendall metric (Figure \ref{fig:rejectrates}, center) suggested that
the monotonicity property of the first- and last-stage item probabilities is not specific of the EPL, but it is shared by other rankings models too. This implies that \eqref{t:T} does not discriminate the EPL from other parametric families with the suitable flexibility to describe an underlying stagewise elicitation process with a certain coherence, over the stages, about the preferences of the items. This is also the case of some subclasses of the Thurstone model. In fact, besides the trivial case of the ordered statistics model with Gumbel distributions for the latent utilities (corresponding to the PL), the property is recovered also when adopting independent normals with varying means and constant variances (the so-called Case V model). Conversely, the property does not hold in general for the distance-based models with any metric other than the Kendall one. Although an exact computation of the marginal item distributions in each position is a difficult task, for a specific ranking model these claims can be easily verified via a simulation approach. 

Another stable evidence highlighted by the comparative analysis concerns the best-performing diagnostic, which turned to be the one relying on the marginal item distributions. However, it is no less apparent that, for higher values of $K$ and $N$, the performance of the new IIA statistic is pretty much equivalent to that of the chi-squared based on the marginal distributions and, in general, always better than the remaining competing statistics typically used in the real-data applications. Moreover, the computation of $X^2_{\text{IIA}}$ is remarkably less computational demanding than $X^2_{\text{M}}$.

\section{Likelihood-free estimation of the reference order}
\label{s:euristic}
Rather than from a model specification perspective, in this section we explored the utility of the statistic \eqref{t:T} from the inferential point of view.

\subsection{The novel heuristic method}
\label{ss:noveuristic}
Let $\mathbf{T}(\upi)=(T_{jj'}(\upi))$ be the $K\times K$ matrix with entries defined in \eqref{t:Tentry}. The computation of $\mathbf{T}(\upi)$ is illustrated with an example reported in Appendix B. For each component $T_{jj'}(\upi)$, the following inequality holds
\begin{equation*}
T_{jj'}(\upi)\leq u_K,
\end{equation*}
where the upper bound corresponds to the constant value in the main diagonal of the matrix, that is,
%
\begin{equation*}
\begin{split}
u_K=T_{jj}(\upi)=\sum_{l=1}^K\lvert2l-(K+1)\rvert=2\left(\sum_{l=1}^{(K-1)/2}2l\right)^{K\,\text{mod}\,2}\left(\sum_{l=0}^{K/2-1}(2l+1)\right)^{1-K\,\text{mod}\,2}.
\end{split}
\end{equation*}
This means that the maximum value in $\mathbf{T}(\upi)$ depends on data only through $K$: for $K$ odd, $u_K$ is the double sum of the first $(K+1)/2$ even numbers (starting from zero); conversely, for $K$ even, $u_K$ is the double sum of the first $K/2$ even numbers.

Our heuristic method to estimate the unknown parameter $\rho$ is composed of the following steps:
\begin{enumerate}
\item compute
$$\mathbf{D}(\upi)=\lvert\mathbf{T}(\upi)-u_k\mathbf{J}_K\rvert,$$
where $\mathbf{J}_K$ is $K\times K$ all-ones matrix, so that each component $D_{jj'}(\upi)$ can be interpreted as a measure of the distance between positions $j$ and $j'$ in the sequential rank assignment process; 
\item use the matrix $\mathbf{D}(\upi)$ as the input of a Principal Component Analysis (PCA);
\item estimate $\rho$ by taking the non-decreasing ordering of the scores $(\zeta_1,\dots,\zeta_K)$ of the $K$ positions on the first PC,
given by
\begin{equation*}
\hat \rho=(\hat \rho(1),\dots,\hat \rho(K)):\quad \zeta_{\hat \rho(1)}\leq\cdots\leq \zeta_{\hat \rho(K)}.
\end{equation*}
\end{enumerate}

\subsection{Simulation study on the heuristic method}
\label{ss:applheur}

The inferential effectiveness of the proposal to recover the true discrete parameter was explored by means of a simulation study with a varying cardinality $K$ of the item set and sample size $N$. For each possible combination $(K,N)$, where $K \in \{5,10,15 \}$ and $N \in \{50,200,1000,10000\}$, we drew 100 datasets $\underline\pi^{-1}_{(R)}$ with $R=1,\dots,100$ from the EPL according to the following scheme
\begin{eqnarray*}
\dot\rho^{(R)} & \sim & \Unif\left\{\mathcal{S}_K\right\},\\
\dot{p}_i^{(R)} & \overset{\text{iid}}{\sim} & \Unif(0,1)\qquad i=1,\dots,K,\\
\underline\pi^{-1}_{(R)}| \dot\rho^{(R)}, \underline{\dot{p}}^{(R)}& \sim & \text{EPL}(\dot\rho^{(R)},\underline{\dot{p}}^{(R)}).
\end{eqnarray*}
%
%
%
For comparison purposes, we inferred the reference order of each simulated sample $\underline\pi^{-1}_{(R)}$ with the heuristic strategy described above, with the one replacing the PCA with the Multidimensional Scaling (MDS) and with the MLE approach via the EMM algorithm \citep{Mollica:Tardella}, which is considered as the reference method for the present estimation task.
Finally, the estimation performance of the competing strategies was compared in terms of:
\begin{itemize}
      \item[-] \% recoveries = $ \sum_{R=1}^{100}  I_{[\dot\rho^{(R)}=\hat{\rho}^{(R)}]}$: percentage of matching between the estimated and the actual reference order; 
      \item[-] $\bar{r}_{\text{Spear}}(\dot\rho,\hat{\rho}) = \frac{1}{100} \sum_{R=1}^{100}  r_{\text{Spear}}(\dot\rho^{(R)},\hat{\rho}^{(R)})$: average cograduation between the estimated and the actual reference order computed with Spearman's rank correlation coefficient. 
\end{itemize}

Comparative results are shown in Table \ref{t:simul}.
\begin{table}[t]
\small
\caption{Inferential performance of the heuristic methods via PCA and MDS to estimate the reference order on simulated data compared to the MLE via EMM algorithm.}
 \label{t:simul}
 \centering
\begin{tabular}{lccccccc}
& \multicolumn{3}{c}{$\%$ recoveries} & &
 \multicolumn{3}{c}{$\bar{r}_{\text{Spear}}(\dot\rho,\hat{\rho})$} \\
\cline{2-4}
\cline{6-8}
$(K,N)$  & PCA & MDS &  MLE & & PCA & MDS & MLE\\
\hline
$(5,50)$       & 58 & 60 & 45     & & 0.86 & 0.86 & 0.69\\
$(5,200)$     & 79 & 80 & 77     & & 0.89 & 0.89 & 0.95\\
$(5,1000)$   & 91 & 90 & 100  & & 0.92 & 0.93 & 1.00\\
$(5,10000)$ & 97 & 97 &         & & 0.94 & 0.94 & \\
\hline
$(10,50)$       & 3   & 5 & 3      & & 0.90 & 0.89 & 0.87\\
$(10,200)$     & 14 & 16 & 23  & & 0.93 & 0.93 & 0.98\\
$(10,1000)$   & 55 & 54 & 68  & & 0.96 & 0.95 & 0.99\\
$(10,10000)$ & 79 & 78 &       & & 0.96 & 0.96 & \\
\hline
$(15,50)$       & 0  & 0 & 0      & & 0.92 & 0.92 & 0.91\\
$(15,200)$     & 0  & 1 & 3      & & 0.94 & 0.94 & 0.97\\
$(15,1000)$   &13 & 15 & 26  & & 0.97 & 0.97 & 0.99\\
$(15,10000)$ & 48 & 51 &      & &  0.97 & 0.97 &\\
\hline
\end{tabular}
\end{table}
As apparent, PCA and MDS exhibited essentially the same ability. Compared with the MLE, one can appreciate very good results for the heuristic methods.
The percentage of matching consistently grows with $N$ and, by checking also the cases where there is not an exact correspondence, on average an analogous trend is highlighted for the relative
Spearman correlation.
Additionally, if we look at a fixed $N$,
the percentage of recoveries shows a worse tendency for larger values of $K$. In this regard, the cases $K\in\{10,15\}$ combined with a relatively very low ($N=50$) and very high ($N=10000$) sample size deserve some considerations to stress typical issues which can be encountered in a ranking data analysis. First, in a sparse data situation, all of the estimation techniques exhibit a great uncertainty in recovering the actual $\rho$, testified by the negligible values of the recovery percentage. On the other hand, although a better behavior of the MLE is expected for $N=10000$, this has not been implemented since, without a specialized program, fitting the EPL to a large sample can be deeply computational demanding if not actually unfeasible. This is especially true for the more flexible EPL class, due to the impact of the reference order on the normalizing term of the likelihood and the need of its iterative update during the optimization procedure. Moreover, the computational burden is further aggravated by the multiple initialization practice needed to address the issue of local maxima. Of course, even if one increases the number of starting values, the fast-growing dimension of the reference order space $\mathcal{S}_K$ makes the multiple starting point strategy rapidly ineffective for larger values of $K$ to reach the global mixumum. In the light of these remarks on the MLE, the likelihood-free approach can be motivated as a straightforward method that can be combined with the MLE procedure or with an MCMC method in the Bayesian estimation framework. In fact, without computational costs, it can implemented as a preliminary step of the inferential process to obtain a promising initialization, that can guide the parameter space exploration towards the global optimum and substantially reduce the elaboration time.

\section{Conclusions}
\label{s:conc}

In this work we contributed with new methods for improving the analysis of ranking data under the assumption that the observations were generated from the stagewise EPL distribution. In particular, we focussed on the
gap of specific goodness-of-fit statistics for multistage ranking models
and on the peculiar issue related to the EPL, that concerns the inference on the discrete parameter component.

Inspired by two formal properties of the EPL parametric class, the former discussed and proven for the first time in the present work and the latter inherited from the PL subclass (\textit{independence of irrelevant alternatives}), we constructed and explored the usefulness of two novel sample statistics
to test the appropriateness of the EPL distribution.
The comparative performance of the two diagnostics with respect to more general goodness-of-fit tests for ranking models has been evaluated by means of a simulation study under alternative data-generating models.
On one hand, the comparison highlighted the limits of the statistics based on the property of inverse monotonicity of the last-stage item probabilities to discriminate the EPL from the other distributions. On the other hand, the simulation study identified the generic test statistic based on the first-order marginals as the best-performing one although, for larger values of $K$ and $N$, the proposed IIA diagnostic exhibited an equivalent power. However, we stress that, differently from $X^2_{\text{M}}$, the novel $X^2_{\text{IIA}}$ represents a specific test for the EPL assumption and does not require intensive MC approximations for its computation. To our opinion, the higher power of the two statistics could depend on a better account of
the $K$-dimensional ranking process, that is, the ability of the two statistics $X^2_{\text{IIA}}$ and $X^2_{\text{M}}$
to span the whole multivariate dependence structure, rather than only univariate or bivariate marginal features of the preference elicitation, such as the tests based on the top position frequencies or on the PC. In this sense, the originality of the simulation results under the EPL specification could stimulate future research on the critical issue concerning the evaluation of the adequacy of ranking models. 

Then, we
revised
the usefulness of the property of inverse monotonicity of the last-stage item probabilities from the inferential perspective, as the core ingredient of a heuristic method to estimate $\rho$.
It is aimed at addressing the estimation issue
with lower computational costs, by returning a promising sample-based evaluation of $\rho$ that can be used as a good initialization of iterative inferential procedures. Hence, the new likelihood-free strategy could fruitfully replace the more conventional and time-cunsuming multiple-initialization practice to attain the global optimum of the likelihood. The utility of the proposal has been checked with a comparative simulation study. 

As a possible future development, we would like to contribute with the introduction and evaluation of other specific goodness-of-fit tests for the class of stagewise models, in order to gain further improvement over standard ranking model diagnostics.
Finally, another valuable direction of research could be the Bayesian extension of the novel diagnostic tools allowing for model adequacy evaluation via posterior predictive checks.

\appendix
\section*{Appendix A: Formal proof}
Without loss of generality, let $\text{EPL}(\rho,\underline{p})$ with $p_1\leq\cdots\leq p_K$ be the data generating mechanism. We are interested in determining the ordering of the probability masses relative to the marginal item distribution at the last stage (stage $K$). To simplify the notation of the proof, we consider the following conventions: for $t=1,\dots,K$, we set $i_t=\eta^{-1}(t)=\pi^{-1}(\rho(t))$ to denote the label of the item selected at stage $t$ and with $\underline{p}_{[D]}=\sum_{i\in D}p_i$ the sum of the support parameters of the items belonging to the choice set $D\subseteq I$. Let us write the marginal probability for item 1 to be chosen in the final step $K$ of the ranking process. This can be obtained by marginalizing out the entries of 
the previous $K-1$ stages, that is,
\begin{equation*}
\begin{split}
q^{[K]}_{1}=\mathbf{P}&_\text{EPL}(i_K=1|\rho,\underline{p})=\mathbf{P}_\text{PL}(\pi^{-1}(\rho(K))=1|\underline{p})
=\mathbf{P}_\text{PL}(\eta^{-1}(K)=1|\underline{p})=\mathbf{P}_\text{PL}(i_K=1|\underline{p})=\\
&=\sum_{i_1\in I\setminus\{1\}}
\cdots
\sum_{i_t\in I\setminus\{1,i_1,\dots,i_{t-1}\}}
\cdots
\sum_{i_{K-1}\in I\setminus\{1,i_1,\dots,i_{K-2}\}}\mathbf{P}_\text{PL}(i_1,\dots,i_{t},\dots,i_{K-1},i_K=1|\underline{p})\\
&=\sum_{i_1\in I\setminus\{1\}}\mathbf{P}_\text{PL}(i_1|\underline{p})
\times\cdots\times
\sum_{i_t\in I\setminus\{1,i_1,\dots,i_{t-1}\}}\mathbf{P}_\text{PL}(i_t|i_1,\dots,i_{t-1},\underline{p})
\times\cdots\times\\
&\times\cdots\times
\sum_{i_{K-1}\in I\setminus\{1,i_1,\dots,i_{K-2}\}}\mathbf{P}_\text{PL}(i_{K-1}|i_1,\dots,i_{K-2},\underline{p})\\
&=\sum_{i_1\in I\setminus\{1\}}\frac{p_{i_1}}{\underline{p}_{[I]}}
\times\cdots\times
\sum_{i_t\in I\setminus\{1,i_1,\dots,i_{t-1}\}}\frac{p_{i_t}}{\underline{p}_{[I\setminus \{i_1,\dots,i_{t-1}\}]}}
\times\cdots\times
\sum_{i_{K-1}\in I\setminus\{1,i_1,\dots,i_{K-2}\}}\frac{p_{i_{K-1}}}{\underline{p}_{[I\setminus \{i_1,\dots,i_{K-2}\}]}}\\
&=\sum_{i_1\in I\setminus\{1\}}
\cdots
\sum_{i_t\in I\setminus\{1,i_1,\dots,i_{t-1}\}}
\cdots
\sum_{i_{K-1}\in I\setminus\{1,i_1,\dots,i_{K-2}\}}
\frac{p_{i_1}\cdots p_{i_t}\cdots p_{i_{K-1}}}
{\underline{p}_{[I]}\cdots\underline{p}_{[I\setminus \{i_1,\dots,i_{t-1}\}]}\cdots
\underline{p}_{[I\setminus \{i_1,\dots,i_{K-2}\}]}} \\
&=\sum_{i_1\in I\setminus\{1\}}
\cdots
\sum_{i_t\in I\setminus\{1,i_1,\dots,i_{t-1}\}}
\cdots
\sum_{i_{K-1}\in I\setminus\{1,i_1,\dots,i_{K-2}\}}
\frac{p_{i_1}\cdots p_{i_t}\cdots p_{i_{K-1}} p_1}
{\underline{p}_{[I]}\cdots\underline{p}_{[I\setminus \{i_1,\dots,i_{t-1}\}]}\cdots
\underline{p}_{[I\setminus \{i_1,\dots,i_{K-2}\}] }p_1}.
\end{split}
\end{equation*}
The analogous marginal probability corresponding to the selection of an item $j\neq1$ at stage $K$ is
\begin{equation*}
\begin{split}
q^{[K]}_{j}=\mathbf{P}&_\text{EPL}(i_K=j|\rho,\underline{p})=\mathbf{P}_\text{PL}(\pi^{-1}(\rho(K))=j|\underline{p})
=\mathbf{P}_\text{PL}(\eta^{-1}(K)=j|\underline{p})=\mathbf{P}_\text{PL}(i_K=j|\underline{p})=\\
&=\sum_{i_1\in I\setminus\{j\}}
\cdots
\sum_{i_t\in I\setminus\{j,i_1,\dots,i_{t-1}\}}
\cdots
\sum_{i_{K-1}\in I\setminus\{j,i_1,\dots,i_{K-2}\}}
\frac{p_{i_1}\cdots p_{i_t}\cdots p_{i_{K-1}}}
{\underline{p}_{[I]}\cdots\underline{p}_{[I\setminus \{i_1,\dots,i_{t-1}\}]}\cdots
\underline{p}_{[I\setminus \{i_1,\dots,i_{K-2}\}]}}\\
&=\sum_{i_1\in I\setminus\{1\}}
\cdots
\sum_{i_t\in I\setminus\{1,i_1,\dots,i_{t-1}\}}
\cdots
\sum_{i_{K-1}\in I\setminus\{1,i_1,\dots,i_{K-2}\}}
\frac{p_{i_1}\cdots p_{i_t}\cdots p_{i_{K-1}} p_j}
{\underline{p}_{[I]}\cdots\underline{p}_{[I\setminus \{i_1,\dots,i_{t-1}\}]}\cdots
\underline{p}_{[I\setminus \{i_1,\dots,i_{K-2}\}] }p_j}.
\end{split}
\end{equation*}

The $(K-1)!$ ratios in both masses correspond to all possible first $K-1$ stage sampling sequences. We remind that the full stage sampling sequences $(i_1,i_2,...,i_t,...,i_{K-1},i_K)$ in the two masses end with 1 and $j$ respectively. Note that all the $(K-1)!$ ratios that are summed in both expressions have been multiplied respectively by $p_1/p_1$ and $p_j/p_j$, so that all the numerators are made by the same quantity.
This simplifies the comparison between $q^{[K]}_{1}$ and $q^{[K]}_{j}$ since the numerator is always equal to the product of $\textit{all}$ the support parameter components hence, if we want to assess the relative magnitude of the two probabilities,
we should concentrate on the relative magnitude of the
$(K-1)!$ denominators,
which are all made of $K$ factors.

We can look more specifically to the denominators and revisit their notation, in order to simplify the comparison task. In the
denominators of $q^{[K]}_{1}$, the $K$-tuple of indeces $(i_1,i_2,...,i_t,...,i_{K-1},i_K)$ is such that the last entry $i_K $ is fixed ($i_K=1$), whereas the first $K-1$ entries range over the set of permutations of the remaining integers in $A=I\setminus{\{1\}}$. We can list the set of permutations of  $K$-tuple of indeces $(i_1,i_2,...,i_t,...,i_{K-1},1)$
by using $(a_{\sigma},1)$ with $a=(2,3,\dots,K)$, 
$a_{\sigma} = (a_{\sigma(1)},...,a_{\sigma(K-1)})$
and $\sigma \in {\cal S}_{K-1}$. Similarly, in the denominators of $q^{[K]}_{j}$, the last component of the stage sampling sequence is fixed ($i_K=j$), whereas the first $K-1$ entries range over the permutations of the integers in $B=I\setminus{\{j\}}$. 
Analogously, we can list the set of permutations of  $K$-tuple of indeces $(i_1,i_2,...,i_t,...,i_{K-1},j)$
by using $(b_{\sigma},j)$ with $b=(2,3,...,j-1,1,j+1,...,K)$, $b_{\sigma} = (b_{\sigma(1)},...,b_{\sigma(K-1)})$.
In so doing, we can make a one-to-one comparison of all the homologous denominators respectively indexed by $(a_{\sigma},1)$ and $(b_{\sigma},j)$ by using the same $\sigma \in {\cal S}_{K-1}$. We remark that, in this way, the $\sigma(t)$-th component of $a_\sigma$ coincides with the 
$\sigma(t)$-th component of $b_{\sigma}$, the only exception being $\sigma(t^*)=j-1$ for which $a_{\sigma(t^*)}=j$ and $b_{\sigma(t^*)}=1$. 
In order to rewrite the $K$ factors in the denominators, we will use the subsets 
$A^\sigma_t$ and $B^\sigma_t$ representing the item subsets which comprise, regardless of their order, the components $(a_{\sigma(t)},\dots,a_{\sigma(K-1)},1)$ and $(b_{\sigma(t)},\dots,b_{\sigma(K-1)},j)$ respectively. Hence, the homologous denominators to be compared will be written as $\underline{p}_{[A^\sigma_1]}\times\cdots\times\underline{p}_{[A^\sigma_t]}\times\cdots\times\underline{p}_{[A^\sigma_K]}$ and $\underline{p}_{[B^\sigma_1]}\times\cdots\times\underline{p}_{[B^\sigma_t]}\times\cdots\times\underline{p}_{[B^\sigma_K]}$.
Now we observe that $A^\sigma_1=B^\sigma_1=I$, hence the first factor $\underline{p}_{[A^\sigma_1]}=\underline{p}_{[B^\sigma_1]}=\underline{p}_{[I]}$ is equal to the sum of all the support parameter components. For similar arguments and in the light of the previous remark, we can claim that
$\underline{p}_{[A^\sigma_t]}=\underline{p}_{[B^\sigma_t]}$ also for $t\leq t^*$. For all $t>t^*$, we have that $A^\sigma_t$ differs from $B^\sigma_t$ since, by construction, the former always contains item 1 and not item $j$ while the latter always contains item $j$ but not item 1, as apparent from comparing $(a_{\sigma(t)},\dots,a_{\sigma(K-1)},1)$ and $(b_{\sigma(t)},\dots,b_{\sigma(K-1)},j)$. Hence, the sums $\underline{p}_{[A^\sigma_t]}$ and $\underline{p}_{[B^\sigma_t]}$ differ only for the presence of $p_1$ in the former, replaced by the presence of $p_j$ in the latter, which implies $\underline{p}_{[A^\sigma_t]}\leq\underline{p}_{[B^\sigma_t]}$. Finally, for all the $K$ factors, we have $\underline{p}_{[A^\sigma_t]}\leq\underline{p}_{[B^\sigma_t]}$ for all $t$ and $\sigma$, hence the opposite inequality holds for the sum of the reciprocals, yielding 
$q^{[K]}_{1}\geq q^{[K]}_{j}$ for $j\neq 1$.
The same argument can be extended for each item $i$ such that $p_i\leq p_j$, leading to 
$$q^{[K]}_{1}\geq\cdots\geq q^{[K]}_{K},$$
i.e., the probability masses of the marginal item distribution at the last stage follow the reverse order of the support parameters. The trick of the proof is the definition of a one-to-one mapping between the $(K-1)!$ terms of the two sums $q^{[K]}_{1}$ and $q^{[K]}_{j}$, obtained by matching the selection stage of the $j$-th item in the sequence $(i_1,i_2,...,i_t,...,i_{K-1},1)$ with the selection stage of item 1 in the sequence $(i_1,i_2,...,i_t,...,i_{K-1},j)$ and all the other item selections in the first $K-1$.

Each marginal probability is the result of the sum of $(K-1)!$ ratios due to all possible configurations of the first $K-1$ entries of $\eta^{-1}$. In the comparison between the two masses, one can note that the numerators of the ratios implicated in the former probability are greater than/equal to those of the latter probability, since they never involve the minimum support parameter $p_1$. For the comparison of the denominators, one can consider the one-to-one mapping between the $(K-1)!$ terms of the two sums obtained by matching the selection stage of the $j$-th item in the first probability with the selection stage of item 1 in the latter probability.
It follows that the denominators of the ratios implicated in the former probability are lower than/equal to those of the latter probability, since they always involve the minimum support parameter $p_1$. This implies $q^{[K]}_{1}=\mathbf{P}_\text{EPL}(i_K=1|\rho,\underline{p})\geq \mathbf{P}_\text{EPL}(i_K=j|\rho,\underline{p})=q^{[K]}_{j}$ for $j\neq 1$. Iterating the same argument for each item leads to 
$$\mathbf{P}_\text{EPL}(i_K=1|\rho,\underline{p})\geq\cdots\geq\mathbf{P}_\text{EPL}(i_K=K|\rho,\underline{p}),$$
i.e., the probability masses of the marginal item distribution at the last stage follow the reverse order of the support parameters. 

Conversely, the probability masses of the marginal item distribution at the first stage follow the same order of the support parameters. It can be easily proven by observing that
\begin{equation*}
\mathbf{P}_\text{EPL}(i_1=j|\rho,\underline{p})
=\mathbf{P}_\text{PL}(i_1=j|\underline{p})=\frac{p_j}{\underline{p}_{[I]}}\propto p_j\qquad j=1,\dots,K.
\end{equation*}

\section*{Appendix B: an example of matrix $\mathbf{T}(\upi)$ under the EPL specification}
By using the \texttt{rPLMIX} function of the \texttt{R} package \texttt{PLMIX} \citep{MollicaPLMIX}, one can simulate $N=100$ orderings of $K=5$ items from a genuine EPL model, with a parameter configuration given by
$$\rho=(1,5,2,4,3)\quad\text\quad\underline{p}=(0.15,0.4,0.12,0.08,0.25).$$
Under the above EPL specification, the expected rankings of the items in order of occurrence at the first and the last stage are indicated in the two rows of Table~\ref{t:firstlast}.
\begin{table}[t]
\caption{Expected rankings of the items in terms of number of selections at the first and the last stage for an $\text{EPL}(\rho=(1,5,2,4,3),\underline{p}=(0.15,0.4,0.12,0.08,0.25))$. The true first and last stage ranks correspond, respectively, to rank 1 and 3.}
\label{t:firstlast}
\centering
\begin{tabular}{cccccc}
 & \multicolumn{5}{c}{Item} \\ 
\cline{2-6}  
 & 1 & 2 & 3 & 4 & 5 \\ 
\cline{2-6}  
$\text{rank}\left(\underline{r}^{[1]}_{1}\right)$ & 3 & 1 & 4 & 5 & 2  \\
$\text{rank}\left(\underline{r}^{[K]}_{3}\right)$  & 3 & 5 & 2 & 1 & 4  \\
\hline
Sum of ranks  & 6 & 6 & 6 & 6 & 6  \\
\end{tabular}
\end{table}
The matrix $\mathbf{T}(\upi)=(T_{jj'}(\upi))$ for all pairs $j,j'=1,\dots,K$ is shown in Table~\ref{t:Tmatrix}. 
\begin{table}[t]
\caption{Matrix $\mathbf{T}(\upi)$ for a simulated sample from the $\text{EPL}(\rho=(1,5,2,4,3),\underline{p}=(0.15,0.4,0.12,0.08,0.25))$. The true first and last stage ranks correspond respectively to rank 1 and 3, yielding the position of the minimum entries of the matrix (in bold).}
\label{t:Tmatrix}
\centering
\begin{tabular}{cccccc}
 & \multicolumn{5}{c}{$j'$} \\ 
\cline{2-6}  
$j$ & 1 & 2 & \bf{3} & 4 & 5 \\ 
\hline
\bf{1} & 12 & 12 & \bf{1} & 6 & 10  \\
2  & 12 & 12 & 2 & 4 & 11  \\
3 & 1 & 2 & 12 & 9 & 5  \\
4  & 6 & 4 & 9 & 12 & 10  \\
5 & 10 & 11 & 5 & 10 & 12  \\
\hline
\end{tabular}
\end{table}
The true first and last stage ranks correspond respectively to rank 1 and 3, yielding the observed value of the EPL statistic $T_m(\upi)=1$, which is actually the global minimum of the whole matrix $\mathbf{T}(\upi)$ in correspondence of the pair $(j,j')=(1,3)$ of the true first and last stage ranks.

\end{document}